\documentclass[11pt]{article}

\usepackage{fullpage,amsmath,amsfonts,amsthm,mathrsfs,xspace}
\usepackage{times,mathtime}

\makeatletter
\def\@thm#1#2#3{%
  \ifhmode\unskip\unskip\par\fi
  \normalfont
  \trivlist
  \let\thmheadnl\relax
  \let\thm@swap\@gobble
  \thm@notefont{\bfseries\upshape}
  \thm@headpunct{.}
  \thm@headsep 5\p@ plus\p@ minus\p@\relax
  \thm@space@setup
  #1
  \@topsep \thm@preskip               
  \@topsepadd \thm@postskip           
  \def\@tempa{#2}\ifx\@empty\@tempa
    \def\@tempa{\@oparg{\@begintheorem{#3}{}}[]}%
  \else
    \refstepcounter{#2}%
    \def\@tempa{\@oparg{\@begintheorem{#3}{\csname the#2\endcsname}}[]}%
  \fi
  \@tempa
}
\makeatother

\newtheorem{prop}{Proposition}

\newtheorem{theorem}[prop]{Theorem}
\newtheorem{fact}[prop]{Fact}

\newtheorem{lemma}[prop]{Lemma}
\newtheorem{conjecture}[prop]{Conjecture}
\newtheorem{corollary}[prop]{Corollary}
\newtheorem{claim}[prop]{Claim}
\theoremstyle{definition}
\newtheorem{definition}{Definition}

\newtheorem*{remark}{Remark}

\newcommand{\mypar}[1]{\vspace{-2ex}\paragraph{#1}}

\newcommand{\cB}{\mathcal{B}}
\renewcommand{\b}{\{0,1\}}

\newcommand{\cE}{\mathcal{E}}
\newcommand{\eps}{\varepsilon}
\DeclareMathOperator{\E}{\mathbb{E}}
\DeclareMathOperator{\e}{e}

\DeclareMathOperator{\Hyp}{\mathrm{Hyp}}

\newcommand{\cC}{\mathcal{C}}
\newcommand{\cP}{\mathcal{P}}
\newcommand{\cQ}{\mathcal{Q}}

\DeclareMathSymbol{\N}{\mathbin}{AMSb}{"4E}
\DeclareMathSymbol{\R}{\mathbin}{AMSb}{"52}
\DeclareMathSymbol{\Z}{\mathbin}{AMSb}{"5A}
\newcommand{\cU}{\mathcal{U}}

\newcommand{\ceq}{\subseteq}

\newcommand{\dist}{\Delta}

\newcommand{\ghd}{\textsc{ghd}\xspace}

\DeclareMathOperator{\vcd}{VC-dim}

\newcommand{\deq}{\mathrel{:=}}  
\newcommand{\ang}[1]{\langle #1 \rangle}

\newcommand{\orth}{\perp}
\newcommand{\comments}[1]{}
\newcommand{\mnote}[1]{}

\title{%
  A Multi-Round Communication Lower Bound for Gap Hamming and 
  Some Consequences\thanks{Work supported in part by an NSF CAREER Award 
  CCF-0448277 and NSF grant EIA-98-02068.}
}

\author{%
  Joshua Brody\\
  \mbox{}\\
  {\small Department of Computer Science}\\
  {\small Dartmouth College}\\
  {\small Hanover, NH 03755, USA}\\
  {\small jbrody@cs.dartmouth.edu}
  \and 
  Amit Chakrabarti\\
  \mbox{}\\
  {\small Department of Computer Science}\\
  {\small Dartmouth College}\\
  {\small Hanover, NH 03755, USA}\\
  {\small ac@cs.dartmouth.edu}
}

\begin{document}

\maketitle


\begin{abstract}

The Gap-Hamming-Distance problem arose in the context of proving space
lower bounds for a number of key problems in the data stream model. In
this problem, Alice and Bob have to decide whether the Hamming distance
between their $n$-bit input strings is large (i.e., at least $n/2 +
\sqrt n$) or small (i.e., at most $n/2 - \sqrt n$); they do not care if
it is neither large nor small. This $\Theta(\sqrt n)$ gap in the problem
specification is crucial for capturing the approximation allowed to a
data stream algorithm.

Thus far, for randomized communication, an $\Omega(n)$ lower bound on
this problem was known only in the one-way setting. We prove an
$\Omega(n)$ lower bound for randomized protocols that use any constant
number of rounds. 

As a consequence we conclude, for instance, that $\eps$-approximately
counting the number of distinct elements in a data stream requires
$\Omega(1/\eps^2)$ space, even with multiple (a constant number of)
passes over the input stream. This extends earlier one-pass lower
bounds, answering a long-standing open question. We obtain similar
results for approximating the frequency moments and for approximating
the empirical entropy of a data stream.

In the process, we also obtain tight $n - \Theta(\sqrt{n}\log n)$ lower
and upper bounds on the one-way deterministic communication complexity
of the problem. Finally, we give a simple combinatorial proof of an
$\Omega(n)$ lower bound on the one-way randomized communication
complexity.

\end{abstract}


\section{Introduction} \label{sec:intro}

This paper concerns communication complexity, which is a heavily-studied
basic computational model, and is a powerful abstraction useful for
obtaining results in a variety of settings not necessarily involving
communication. To cite but two examples, communication complexity has
been applied to prove lower bounds on circuit depth (see,
e.g.,~\cite{KarchmerW90}) and on query times for static data structures
(see, e.g.,~\cite{MiltersenNSW98,Patrascu08ds}). The basic setup
involves two players, Alice and Bob, each of whom receives an input
string. Their goal is to compute some function of the two strings, using
a protocol that involves exchanging a {\em small} number of bits. When
communication complexity is applied as a lower bound technique --- as it
often is --- one seeks to prove that there does not exist a nontrivial
protocol, i.e., one that communicates only a sublinear number of bits,
for computing the function of interest. Naturally, such a proof is more
challenging when the protocol is allowed to be {\em randomized} and err
with some small probability on each input.

The textbook by Kushilevitz and Nisan~\cite{KushilevitzNisan-book}
provides detailed coverage of the basics of communication complexity,
and of a number of applications, including the two mentioned above. In
this paper, we only recap the most basic notions, in
Section~\ref{sec:prelim}.

Our focus here is on a specific communication problem --- the
Gap-Hamming-Distance problem --- that, to the best of our knowledge, was
first formally studied by Indyk and Woodruff~\cite{IndykW03} in FOCS
2003. They studied the problem in the context of proving space lower
bounds for the Distinct Elements problem in the data stream model. We
shall discuss their application shortly, but let us first define our
communication problem precisely.

\mypar{The Problem.}
In the Gap-Hamming-Distance problem, Alice receives a Boolean string
$x\in\b^n$ and Bob receives $y\in\b^n$. They wish to decide whether $x$
and $y$ are ``close'' or ``far'' in the Hamming sense. That is, they
wish to output $0$ if $\dist(x,y) \le n/2 - \sqrt{n}$ and $1$ if
$\dist(x,y) \ge n/2 + \sqrt{n}$. They do not care about the output if
neither of these conditions holds. Here, $\dist$ denotes Hamming
distance. In the sequel, we shall be interested in a parametrized
version of the problem, where the thresholds are set at $n/2 \pm
c\sqrt{n}$, for some parameter $c\in\R^+$.

\mypar{Our Results.}
While we prove a number of results about the Gap-Hamming-Distance
problem here, there is a clear ``main theorem'' that we wish to
highlight. Technical terms appearing below are defined precisely in
Section~\ref{sec:prelim}.
\begin{theorem}[Main Theorem, Informal] \label{thm:main-informal}
  Suppose a randomized $\frac13$-error protocol solves the
  Gap-Hamming-Distance problem using $k$ rounds of communication. Then,
  at least one message must be $n/2^{O(k^2)}$ bits long. In particular,
  any protocol using a constant number of rounds must communicate
  $\Omega(n)$ bits in some round. In fact, these bounds apply to
  deterministic protocols with low distributional error under the
  uniform distribution.
\end{theorem}

Notice that our lower bound applies to the {\em maximum} message length,
not just the {\em total} length.

At the heart of our proof is a round elimination lemma that lets us
``eliminate'' the first round of communication, in a protocol for the
Gap-Hamming-Distance problem, and thus derive a shorter protocol for an
``easier'' instance of the same problem. By repeatedly applying this
lemma, we eventually eliminate all of the communication. We also make
the problem instances progressively easier, but, if the original
protocol was short enough, at the end we are still left with a
nontrivial problem. The resulting contradiction lower bounds the length
of the original protocol. We note that this underlying ``round
elimination philosophy'' is behind a number of key results in
communication complexity~\cite{MiltersenNSW98,Sen03,ChakrabartiR04,%
AdlerDHP06,Chakrabarti07,ViolaW07,ChakrabartiJP08}.

Besides the above theorem, we also prove tight lower {\em and upper}
bounds of $n - \Theta(\sqrt{n}\log n)$ on the one-way deterministic
communication complexity of Gap-Hamming-Distance. Only $\Omega(n)$ lower
bounds were known before. We also prove an $\Omega(n)$ one-way
randomized communication lower bound. This matches earlier results, but
our proof has the advantage of being purely combinatorial. (We recently
learned that Woodruff~\cite{Woodruff09} had independently discovered a
similar combinatorial proof. We present our proof nevertheless, for
pedagogical value, as it can be seen as a generalization of our
deterministic lower bound proof.)

\mypar{Motivation and Relation to Prior Work.}
We now describe the original motivation for studying the
Gap-Hamming-Distance problem. Later, we discuss the consequences of our
Theorem~\ref{thm:main-informal}. In the data stream model, one wishes to
compute a real-valued function of a massively long input sequence (the
data stream) using very limited space, hopefully sublinear in the input
length. To get interesting results, one almost always needs to allow
randomized approximate algorithms. A key problem in this model, that has
seen much research~\cite{FlajoletM85,AlonMS99,BarYossefJKST02,IndykW03,
Woodruff09}, is the Distinct Elements problem: the goal is to estimate
the number of distinct elements in a stream of $m$ elements (for
simplicity, assume that the elements are drawn from the universe $[m] :=
\{1,2,\ldots,m\}$). 

An interesting solution to this problem would give an nontrivial
tradeoff between the quality of approximation desired as the space
required to achieve it.  The best such result~\cite{BarYossefJKST02}
achieved a multiplicative $(1+\eps)$-approximation using space
$\widetilde{O}(1/\eps^2)$, where the $\widetilde{O}$-notation suppresses
$\log m$ and $\log(1/\eps)$ factors. It also processed the input stream
in a single pass, a very desirable property. Soon afterwards, Indyk and
Woodruff~\cite{IndykW03} gave a matching $\Omega(1/\eps^2)$ space lower
bound for one-pass algorithms for this problem, by a reduction from the
Gap-Hamming-Distance communication problem. In SODA 2004,
Woodruff~\cite{Woodruff04} improved the bound, extending it to the full
possible range of subconstant $\eps$, and also applied it to the more
general problem of estimating frequency moments $F_p := \sum_{i=1}^n
f_i^p$, where $f_i$ is the frequency of element $i$ in the input stream.
A number of other natural data stream problems have similar space lower
bounds via reductions from Gap-Hamming, a more recent example being the
computation of the empirical entropy of a stream~\cite{ChakrabartiCM07}.

The idea behind the reduction is quite simple: Alice and Bob can convert
their Gap-Hamming inputs into suitable streams of integers, and then
simulate a one-pass streaming algorithm using a single round of
communication in which Alice sends Bob the memory contents of the
algorithm after processing her stream. In this way, an $\Omega(n)$
one-way communication lower bound translates into an $\Omega(1/\eps^2)$
one-pass space lower bound.  Much less simple was the proof of the
communication lower bound itself.  Woodruff's proof~\cite{Woodruff04}
required intricate combinatorial arguments and a fair amount of
complex calculations. Jayram et al.~\cite{JayramKS07} later provided a
rather different proof, based on a simple geometric argument, coupled
with a clever reduction from the \textsc{index} problem. A version of
this proof is given in Woodruff's Ph.D. thesis~\cite{Woodruff-thesis}.
In Section~\ref{sec:one-round}, we provide a still simpler direct
combinatorial proof, essentially from first principles.

All of this left open the tantalizing possibility that a second pass
over the input stream could drastically reduce the space required to
approximate the number of distinct elements --- or, more generally, the
frequency moments $F_p$. Perhaps $\widetilde{O}(1/\eps)$ space was
possible? This was a long-standing open problem~\cite{Kumar06-talk} in
data streams.  Yet, some thought about the underlying Gap-Hamming
communication problem suggested that the linear lower bound ought to
hold for general communication protocols, not just for one-way
communication. This prompted the following natural conjecture.

\begin{conjecture} \label{conj:ghd}
  A $\frac13$-error randomized communication protocol for the
  Gap-Hamming-Distance problem must communicate $\Omega(n)$ bits in
  total, irrespective of the number of rounds of communication.
\end{conjecture}

An immediate consequence of the above conjecture is that a second pass
does {\em not} help beat the $\Omega(1/\eps^2)$ space lower bound for
the aforementioned streaming problems; in fact, no constant number of
passes helps. Our Theorem~\ref{thm:main-informal} does {\em not} resolve
Conjecture~\ref{conj:ghd}.  However, it {\em does} imply the
$\Omega(1/\eps^2)$ space lower bound with a constant number of passes.
This is because we {\em do} obtain a linear communication lower bound
with a constant number of rounds.

\mypar{Finer Points.}
To better understand our contribution here, it is worth considering some
finer points of previously known lower bounds on Gap-Hamming-Distance,
including some ``folklore'' results. The earlier one-way $\Omega(n)$
bounds were {\em inherently} one-way, because the \textsc{index} problem
has a trivial two-round protocol. Also, the nature of the reduction
implied a distributional error lower bound for Gap-Hamming only under a
somewhat artificial input distribution. Our bounds here, including our
one-way randomized bound, overcome this problem, as does the recent
one-way bound of Woodruff~\cite{Woodruff09}: they apply to the uniform
distribution. As noted by Woodruff~\cite{Woodruff09}, this has the
desirable consequence of implying space lower bounds for the Distinct
Elements problem under weaker assumptions about the input stream: it
could be random, rather than adversarial.

Intuitively, the uniform distribution is the hard case for the
Gap-Hamming problem. The Hamming distance between two uniformly
distributed $n$-bit strings is likely to be just around the
$n/2\pm\Theta(\sqrt n)$ thresholds, which means that a protocol will
have to work hard to determine which threshold the input is at. Indeed,
this line of thinking suggests an $\Omega(n)$ lower bound for
distributional complexity --- under the uniform distribution --- on the
{\em gapless} version of the problem.  Our proofs here confirm this
intuition, at least for a constant number of rounds.

It is relatively easy to obtain an $\Omega(n)$ lower bound on the {\em
deterministic} multi-round communication complexity of the problem. One
can directly demonstrate that the communication matrix contains no large
monochromatic rectangles (see, e.g.~\cite{Woodruff-thesis}). Indeed, the
argument goes through even with gaps of the form $n/2\pm\Theta(n)$,
rather than $n/2\pm\Theta(\sqrt n)$. It is also easy to obtain an
$\Omega(n)$ bound on the randomized complexity of the gapless problem,
via a reduction from \textsc{disjointness}. Unfortunately, the known
hard distributions for \textsc{disjointness} are far from uniform, and
\textsc{disjointness} is actually very easy under a uniform input
distribution. So, this reduction does not give us the results we want.

Furthermore, straightforward rectangle-based methods
(discrepancy/corruption) fail to effectively lower bound the randomized
communication complexity of our problem. This is because there {\em do}
exist very large near-monochromatic rectangles in its communication
matrix. This can be seen, e.g., by considering all inputs $(x,y)$ with
$x_i = y_i = 0$ for $i \in [n/100]$.

\mypar{Connection to Decision Trees and Quantum Communication.}
We would like to bring up two other illuminating observations. Consider
the following query complexity problem: the input is a string $x\in\b^n$
and the desired output is $1$ if $|x| \ge n/2 + \sqrt n$ and $0$ if $|x|
\le n/2 - \sqrt n$. Here, $|x|$ denotes the Hamming weight of $x$. The
model is a randomized decision tree whose nodes query individual bits of
$x$, and whose leaves give outputs in $\b$. It is not hard to show that
$\Omega(n)$ queries are needed to solve this problem with $\frac13$
error. Essentially, one can do no better than sampling bits of $x$ at
random, and then $\Omega(1/\eps^2)$ samples are necessary to distinguish
a biased coin that shows heads with probability $\frac12 + \eps$ from
one that shows heads with probability $\frac12 - \eps$. 

The Gap-Hamming-Distance problem can be seen as a generalization of this
problem to the communication setting. Certainly, any efficient decision
tree for the query problem implies a correspondingly efficient
communication protocol, with Alice acting as the querier and Bob acting
as the responder (say). Conjecture~\ref{conj:ghd} says that no better
communication protocols are possible for this problem.

This query complexity connection brings up another crucial point. The
{\em quantum} query complexity of the above problem can be shown to be
$O(\sqrt n)$, by the results of Nayak and Wu~\cite{NayakW99}. This in
turn implies an $O(\sqrt n \log n)$ quantum communication protocol for
Gap-Hamming, essentially by carefully ``implementing'' the quantum query
algorithm, as in Razborov~\cite{Razborov02}. Therefore, any technique
that seeks to prove an $\Omega(n)$ lower bound for Gap-Hamming (under
classical communication) must necessarily fail for quantum protocols.
This rules out several recently-developed methods, such as the
factorization norms method of Linial and Shraibman~\cite{LinialS07} and
the pattern matrix method of Sherstov~\cite{Sherstov08}.

\mypar{Connections to Recent Work.}
Our multi-round $\Omega(n)$ bound turns out to also have
applications~\cite{ArackaparambilBC09} to the communication complexity
of several distributed ``functional monitoring'' problems, studied
recently by Cormode et al.~\cite{CormodeMY08} in SODA 2008. Also, our
lower bound approach here uses and extends a subspace-finding technique
recently developed by Brody~\cite{Brody09} to prove lower bounds on
multiparty pointer jumping.


\section{Basic Definitions, Notation and Preliminaries}\label{sec:prelim}

We begin with definitions of our central problem of interest, and
quickly recall some standard definitions from communication complexity.
Along the way, we also introduce some notation that we use in the rest
of the paper.

\begin{definition}
  For strings $x,y\in\b^n$, the Hamming distance between $x$ and $y$,
  denoted $\dist(x,y)$, is defined as the number of coordinates
  $i\in[n]$ such that $x_i \ne y_i$.
\end{definition}

\begin{definition}[Gap-Hamming-Distance problem]
  Suppose $n\in\N$ and $c\in\R^+$.  The $c$-Gap-Hamming-Distance partial
  function, on $n$-bit inputs, is denoted $\ghd_{c,n}$ and is defined as
  follows.
  
  \[ 
    \ghd_{c,n}(x,y) ~=~ \begin{cases}
      1 \, ,     & \> \text{~if~} \dist(x,y) \ge n/2 + c\sqrt{n} \, , \\ 
      0 \, ,     & \> \text{~if~} \dist(x,y) \le n/2 - c\sqrt{n}  \, ,\\ 
      \star \, , & \> \text{~otherwise.}
    \end{cases} 
  \]
  We also use $\ghd_{c,n}$ to denote the corresponding communication
  problem where Alice holds $x\in\b^n$, Bob holds $y\in\b^n$, and the
  goal is for them to communicate and agree on an output bit that
  matches $\ghd_{c,n}(x,y)$.  By convention, $\star$ matches both $0$ 
  and $1$.
\end{definition}

\mypar{Protocols.}
Consider a communication problem $f:\b^n\times\b^n\to\{0,1,\star\}^n$
and a protocol $\cP$ that attempts to solve $f$. We write $\cP(x,y)$ to
denote the output of $\cP$ on input $(x,y)$: note that this may be a
random variable, dependent on the internal coin tosses of $\cP$, if
$\cP$ is a randomized protocol. A deterministic protocol $\cP$ is said
to be correct for $f$ if $\forall\,(x,y):\,\cP(x,y) = f(x,y)$ (the
``$=$'' is to be read as ``matches'').  It is said to have {\em
distributional error} $\eps$ under an input distribution $\rho$ if
$\Pr_{(x,y)\sim\rho}[\cP(x,y) \ne f(x,y)] \le \eps$. A {\em randomized
protocol} $\cP$, using a public random string $r$, is said to be have
error $\eps$ if $\forall\,(x,y):\,\Pr_r[\cP(x,y) \ne f(x,y)] \le \eps$.
A protocol $\cP$ is said to be a {\em $k$-round protocol} if it involves
exactly $k$ messages, with Alice and Bob taking turns to send the
messages; by convention, we usually assume that Alice sends the first
message and the recipient of the last message announces the output. A
$1$-round protocol is also called a {\em one-way protocol}, since the
entire communication happens in the Alice $\to$ Bob direction.

\mypar{Communication Complexity.}
The deterministic communication complexity $D(f)$ of a communication
problem $f$ is defined to be the minimum, over deterministic protocols
$\cP$ for $f$, of the number of bits exchanged by $\cP$ for a worst-case
input $(x,y)$. By suitably varying the class of protocols over which the
minimum is taken, we obtain, e.g., the $\eps$-error randomized, one-way
deterministic, $\eps$-error one-way randomized, and $\eps$-error
$\rho$-distributional deterministic communication complexities of $f$,
denoted $R_\eps(f)$, $D^\to(f)$, $R_\eps^\to(f)$, and
$D_{\rho,\eps}(f)$, respectively. When the error parameter $\eps$ is
dropped, it is tacitly assumed to be $\frac13$; as is well-known, the
precise value of this constant is immaterial for asymptotic bounds.

\begin{definition}[Near-Orthogonality] \label{def:orth}
  We say that strings $x,y\in\b^n$ are $c$-near-orthogonal, and write
  $x\orth_c y$, if $|\dist(x,y) - n/2| < c\sqrt n$. Here, $c$ is a
  positive real quantity, possibly dependent on $n$. Notice that
  $\ghd_{c,n}(x,y) = \star ~\Leftrightarrow~ x \orth_c y$.
\end{definition}

The distribution of the Hamming distance between two uniform random
$n$-bit strings --- equivalently, the distribution of the Hamming weight
of a uniform random $n$-bit string --- is just an unbiased binomial
distribution $\mathrm{Binom}(n,\frac12)$.  We shall use the following
(fairly loose) bounds on the tail of this distribution (see, e.g.,
Feller~\cite{Feller-book}).

\begin{fact} \label{fct:normal-tail}
  Let $T_n(c) = \Pr_x \left[ x \not\orth_c 0^n \right]$, where $x$ is
  distributed uniformly at random in $\b^n$. Let $T(c) =
  \lim_{n\to\infty} T_n(c)$. Then 
  \[
    2^{-3c^2-2} ~\le~ T(c) ~\approx~
    \frac{e^{-2c^2}}{c\sqrt{2\pi}} ~\le~ 2^{-c^2} \, .
  \]
\end{fact}

There are two very natural input distributions for $\ghd_{c,n}$: the
uniform distribution on $\b^n\times\b^n$, and the (non-product)
distribution that is uniform over all inputs for which the output is
precisely defined. We call this latter distribution $\mu_{c,n}$. 

\begin{definition}[Distributions] \label{def:distribs}
  For $n\in\N$, $c\in\R^+$, let $\mu_{c,n}$ denote the uniform
  distribution on the set $\{(x,y)\in\b^n\times\b^n:\, x \not\orth_c
  y\}$. Also, let $\cU_n$ denote the uniform distribution on $\b^n$.
\end{definition}

Using Fact~\ref{fct:normal-tail}, we can show that for a constant $c$
and suitably small $\eps$, the distributional complexities
$D_{\cU_n\times\cU_n,\eps}(\ghd_{c,n})$ and
$D_{\mu_{c,n},\eps}(\ghd_{c,n})$ are within constant factors of each
other. This lets us work with the latter and draw conclusions about the
former. The latter has the advantage that it is meaningful for any $\eps
< \frac12$, whereas the former is only meaningful if $\eps < \frac12
T(c)$. 

Let $\cB(x,r)$ denote the Hamming ball of radius $r$ centered at $x$.
We need use the following bounds on the volume (i.e., size) of a Hamming
ball. Here, $H:[0,1]\to[0,1]$ is the binary entropy function.

\begin{fact}\label{fact:smallball}
  If $r = c\sqrt{n}$, then $(\sqrt{n}/c)^{r} < |\cB(x,r)| < n^r$.
\end{fact}

\begin{fact}\label{fact:largeball}
  If $r = \alpha n$ for some constant $0 < \alpha < 1$, then
  $|\cB(x,r)| \leq 2^{nH(\alpha)}$.
\end{fact}


\section{Main Theorem: Multi-Round Lower Bound}

\subsection{Some Basics}

In order to prove our multi-round lower bound, we need a simple --- yet,
powerful --- combinatorial lemma, known as Sauer's Lemma~\cite{Sauer72}.
For this, we recall the concept of Vapnik-Chervonenkis dimension. Let
$S\ceq\b^n$ and $I\ceq[n]$. We say that $S$ shatters $I$ if the set
obtained by restricting the vectors in $S$ to the coordinates in $I$ has
the maximum possible size, $2^{|I|}$. We define $\vcd(S)$ to be the
maximum $|I|$ such that $S$ shatters $I$.

\begin{lemma}[Sauer's Lemma] \label{lem:sauer}
  Suppose $S\ceq\b^n$ has $\vcd(S) < d$. Then 
  \[
    |S| ~\le~ \sum_{i=0}^d \binom{n}{d} \, .
  \]
\end{lemma}
When $d = \alpha n$ for some constant $\alpha$, then the above sum
can be upper bounded by $2^{nH(\alpha)}$. This yields the following
corollary.
\begin{corollary} \label{cor:vcd}
  If $|S| \ge 2^{nH(\alpha)}$, for a constant $\alpha$, then 
  $\vcd(S) \ge \alpha n$.
\end{corollary}

We now turn to the proof proper. It is based on a round elimination
lemma that serves to eliminate the first round of communication of a
$\ghd$ protocol, yielding a shorter protocol, but for $\ghd$ instances
with weakened parameters. To keep track of all relevant parameters, we
introduce the following notation.
\begin{definition}
  A $[k,n,s,c,\eps]$-protocol is a deterministic $k$-round protocol for
  $\ghd_{c,n}$ that errs on at most an $\eps$ fraction of inputs, under
  the input distribution $\mu_{c,n}$, and in which each message is $s$
  bits long.
\end{definition}

The next lemma gives us the ``end point'' of our round elimination
argument.
\begin{lemma}\label{lem:zero-round}
  There exists no $[0,n,s,c,\eps]$-protocol with $n > 1$, $c =
  o(\sqrt{n})$, and $\eps < \frac12$.
\end{lemma}
\begin{proof}
  With these parameters, $\mu_{c,n}$ has nonempty support. This implies
  $\Pr_{\mu_{c,n}}[\ghd_{c,n}(x,y) = 0] =
  \Pr_{\mu_{c,n}}[\ghd_{c,n}(x,y) = 1] = \frac12$.  Thus, a $0$-round
  deterministic protocol, which must have constant output, cannot
  achieve error less than $\frac12$.
\end{proof}

\subsection{The Round Elimination Lemma}

The next lemma is the heart of our proof. To set up its parameters, we
set $t_0 = (48 \ln 2)\cdot 2^{11k}$, $t = 2^{15k}$, and $b =
T^{-1}(1/8)$, and we define a sequence
$\ang{(n_i,s_i,c_i,\eps_i)}_{i=0}^k$ as follows:
\begin{equation} \label{eq:recurrence}
  \left.\begin{array}{*{2}{r@{~=~}l}}
    n_0    & n \, ,           \qquad & n_{i+1}    & n_i/3 \, , \\
    s_0    & t_0 s \, ,    \qquad & s_{i+1}    & t s_i \, , \\
    c_0    & 10 \, ,          \qquad & c_{i+1}    & 2c_i \, , \\
    \eps_0 & 2^{-2^{11k}} \, , \qquad & \eps_{i+1} & \eps_i/T(c_{i+1}) \, .
  \end{array}\right\}
  \mbox{~for~} 0 \le i < k \, .
\end{equation}

\begin{lemma}[Round Elimination for GHD] \label{lem:round-elim}
  Suppose $0 \le i < k$ and $s_i \le n_i/20$. Suppose there exists 
  a $[k-i,n_i,s_i,c_i,\eps_i]$-protocol. Then there exists 
  a $[k-i-1,n_{i+1},s_{i+1},c_{i+1},\eps_{i+1}]$-protocol.
\end{lemma}
\begin{proof}
  Let $(n,s,c,\eps) = (n_i,s_i,c_i,\eps_i)$ and $(n',s',c',\eps') =
  (n_{i+1},s_{i+1},c_{i+1},\eps_{i+1})$.  Also, let $\mu = \mu_{c,n}$,
  $\mu' = \mu_{c',n'}$, $\ghd = \ghd_{c,n}$ and $\ghd' = \ghd_{c',n'}$.
  %
  Let $\cP$ be a $[k-i,n,s,c,\eps]$-protocol.  Assume, WLOG, that Alice
  sends the first message in $\cP$.  

  Call a string $x_0\in\b^n$ ``good'' if 
  \begin{equation} \label{eq:good}
    \Pr_{(x,y)\sim\mu} [\cP(x,y) \ne \ghd(x,y) \mid x = x_0] ~\le~ 2\eps \, .
  \end{equation}
  By the error guarantee of $\cP$ and Markov's inequality, the number of
  good strings is at least $2^{n-1}$.  There are $2^s \le 2^{n/20}$
  different choices for Alice's first message. Therefore, there is a set
  $M \ceq \b^n$ of good strings such that Alice sends the same first
  message $\mathfrak{m}$ on every input $x \in M$, with $|M| \ge
  2^{n-1-n/20} \ge 2^{nH(1/3)}$. By Corollary~\ref{cor:vcd}, $\vcd(M)
  \ge n/3$. Therefore, there exists a set $I \ceq [n]$, with $|I| = n/3 =
  n'$, that is shattered by $M$.  For strings $x'\in\b^{n'}$ and
  $x''\in\b^{n-n'}$, we write $x'\circ x''$ to denote the string in
  $\b^n$ formed by plugging in the bits of $x'$ and $x''$ (in order)
  into the coordinates in $I$ and $[n]\setminus I$, respectively.

  We now give a suitable $(k-i-1)$-round protocol $\cQ$ for
  $\ghd'$, in which Bob sends the first message. Consider an
  input $(x',y')\in\b^{n'}\times\b^{n'}$, with Alice holding $x'$ and
  Bob holding $y'$.  By definition of shattering, there exists an
  $x''\in\b^{n-n'}$ such that $x := x'\circ x'' \in M$. Alice and Bob
  agree beforehand on a suitable $x$ for each possible $x'$. Suppose Bob
  were to pick a uniform random $y''\in\b^{n-n'}$ and form the string $y :=
  y'\circ y''$. Then, Alice and Bob could simulate $\cP$ on input
  $(x,y)$ using only $k-i-1$ rounds of communication, with Bob starting,
  because Alice's first message in $\cP$ would always be $\mathfrak{m}$.
  Call this randomized protocol $\cQ_1$. We define $\cQ$ to be the
  protocol obtained by running $t$ instances of $\cQ_1$ in parallel,
  using independent random choices of $y''$, and outputting the majority
  answer. Note that the length of each message in $Q$ is $ts = s'$. We
  shall now analyze the error.

  Suppose $x'' \orth_b y''$. Let $d_1 = \dist(x,y) - n/2$, $d_2 =
  \dist(x',y') - n'/2$ and $d_3 = \dist(x'',y'') - (n-n')/2$. Clearly,
  $d_1 = d_2 + d_3$. Also,
  \[
    |d_1| 
    ~\ge~ |d_3| - |d_2|
    ~\ge~ c'\sqrt{n'} - b\sqrt{n-n'}
    ~\ge~ \frac{(c' - b\sqrt2) \sqrt n}{\sqrt3}
    ~\ge~ c\sqrt n \, ,
  \]
  where we used~\eqref{eq:recurrence} and our choice of $b$. Thus, $x
  \not\orth_c y$. The same calculation also shows that $d_1$ and $d_3$
  have the same sign, as $|d_3| > |d_2|$. Therefore $\ghd(x,y) =
  \ghd'(x',y')$. 

  For the rest of the calculations in this proof, fix an input $x'$ for
  Alice, and hence, $x''$ and $x$ as well. For a fixed $y'$, let
  $\cE(y')$ denote the event that $\cP(x,y) \ne \ghd(x,y)$: note that
  $y''$ remains random. Using the above observation (at
  step~\eqref{eq:q1-err-3} below), we can bound the probability that
  $\cQ_1$ errs on input $(x',y')$ as follows. 
  \begin{alignat}{1}
    \Pr_{y} \left[Q_1(x',y') \ne \ghd'(x',y') \mid y'\right]
    &~\le~ \Pr_{y} \left[\cP(x,y) \ne \ghd(x,y) \vee
      \ghd(x,y) \ne \ghd'(x',y') \mid y'\right] \notag \\
    &~\le~ \Pr_{y''} \left[\cE(y')\right]
      + \Pr_{y} \left[\ghd(x,y) \ne \ghd'(x',y') \mid y'\right] \notag \\
    &~\le~ \Pr_{y''} \left[\cE(y')\right] 
      + \Pr_{y''} \left[x'' \not\orth_b y''\right]
      \label{eq:q1-err-3} \\
    &~\le~ \Pr_{y''} \left[\cE(y')\right] + T(b) \notag \\
    &~ = ~ \Pr_{y''} \left[\cE(y')\right] + 1/8 \, ,
      \label{eq:q1-err-5}
  \end{alignat}
  where step~\eqref{eq:q1-err-5} follows from our choice of $b$.  To
  analyze $\cQ$, notice that during the $t$-fold parallel repetition of
  $\cQ_1$, $y'$ remains fixed while $y''$ varies. Thus, it suffices to
  understand how the repetition drives down the sum on the right side
  of~\eqref{eq:q1-err-5}.  Unfortunately, for some values of $y'$, the
  sum may exceed $\frac12$, in which case it will be driven {\em up},
  not down, by the repetition.  To account for this, we shall bound the
  {\em expectation} of the first term of that sum, for a random $y'$.

  To do so, let $z\sim\mu\mid x$ be a random string independent of $y$.
  Notice that $z$ is uniformly distributed on a subset of $\b^n$ of size
  $2^n T(c)$, whereas $y$ is uniformly distributed on a subset of $\b^n$
  of size $2^n T(c')$. (We are now thinking of $x$ as being fixed and
  both $y'$ and $y''$ as being random.) Therefore,
  \begin{alignat}{1}
    \E_{y'} \left[ \Pr_{y''} \left[\cE(y')\right] \right]
    ~ = ~ \Pr_{y} \left[\cE(y')\right]
    &~ = ~ \Pr_{y} \left[\cP(x,y) \ne \ghd(x,y)\right] \notag \\
    &~\le~ \Pr_{z} \left[\cP(x,z) \ne \ghd(x,z)\right] \cdot T(c)/T(c') 
      \notag \\
    &~\le~ 2\eps T(c)/T(c') \, , \label{eq:goodness}
  \end{alignat}
  where~\eqref{eq:goodness} holds because $x$, being good,
  satisfies~\eqref{eq:good}.  Thus, by Markov's inequality,
  \begin{equation}
    \Pr_{y'}\left[ \Pr_{y''}\left[\cE(y')\right] \ge \frac18 \right]
    ~\le~ 16\eps T(c)/T(c') \, . \label{eq:yprime-bad}
  \end{equation}
  If, for a particular $y'$, the {\em bad event} $\Pr_{y''} [\cE(y')]
  \ge \frac18$ does {\em not} occur, then the right side
  of~\eqref{eq:q1-err-5} is at most $1/8 + 1/8 = 1/4$. In other words,
  $\cQ_1$ errs with probability at most $1/4$ for this $y'$. By standard
  Chernoff bounds, the $t$-fold repetition in $\cQ$ drives this error
  down to $(e/4)^{t/4} \le 2^{-t/10} \le \eps_0 \le \eps$. 
  Combining this with~\eqref{eq:yprime-bad}, which bounds the
  probability of the bad event, we get
  \[
    \Pr_{y',r} \left[ \cQ(x',y') \ne \ghd'(x',y') \right]
    ~\le~ 16\eps T(c)/T(c') + \eps 
    ~\le~ \eps/T(c') 
    ~ = ~ \eps' \, ,
  \]
  where $r$ denotes the internal random string of $\cQ$ (i.e., the
  collection of $y''$s used).

  Note that this error bound holds for {\em every} fixed $x'$, and thus,
  when $(x',y')\sim\mu'$. Therefore, we can fix Bob's random coin tosses
  in $\cQ$ to get the desired $[k-i-1,n',s',c',\eps']$-protocol.
\end{proof}

\subsection{The Lower Bound}

Having established our round elimination lemma, we obtain our lower
bound in a straightforward fashion.

\begin{theorem}[Multi-round Lower Bound]
  Let $\cP$ be a $k$-round $\frac13$-error randomized communication
  protocol for $\ghd_{c,n}$, with $c = O(1)$, in which each message is
  $s$ bits long. Then
  \[
    s ~\ge~ \frac{n}{2^{O(k^2)}} \, .
  \]
\end{theorem}
\begin{remark}
  This is a formal restatement of Theorem~\ref{thm:main-informal}.
\end{remark}
\begin{proof}
  For simplicity, assume $c \le c_0 = 10$. Our proof easily applies to a
  general $c = O(1)$ by a suitable modification of the parameters
  in~\eqref{eq:recurrence}. Also, assume $n \ge 2^{4k^2}$, for otherwise
  there is nothing to prove. 

  By repeating $\cP$ $(48 \ln 2)\cdot 2^{11k}=t_0$ times, in parallel,
  and outputting the majority of the answers, we can reduce the error to
  $2^{-2^{11k}} = \eps_0$.  The size of each message is now $t_0s =
  s_0$. Fixing the random coins of the resulting protocol gives us a
  $[k,n_0,s_0,c_0,\eps_0]$-protocol $\cP_0$.

  Suppose $s_i \le n_i/20$ for all $i$, with $0 \le i < k$. We then
  repeatedly apply Lemma~\ref{lem:round-elim} $k$ times, starting with
  $\cP_0$. Eventually, we end up with a
  $[0,n_k,s_k,c_k,\eps_k]$-protocol. Examining~\eqref{eq:recurrence},
  we see that $n_k = n/3^k$, $s_k = 2^{15k^2}s_0 = (48 \ln 2)2^{15k^2 + 11k}s$, and
  $c_k = 10\cdot 2^k$.  Notice that $n_k \ge 2^{4k^2}/3^k > 1$ and
  $c_k = o(\sqrt{n_k})$. We also see that $\ang{c_i}_{i=1}^k$ is an
  increasing sequence, whence $\eps_{i+1}/\eps_i = 1/T(c_{i+1}) \le
  1/T(c_k) \le 2^{3{c_k}^2+2}$, where the final step uses
  Fact~\ref{fct:normal-tail}. Thus,
  \[
    \eps_k 
    ~\le~ \eps_0\big(2^{3c_k^2+2}\big)^k  
    ~ = ~ 2^{-2^{11k}} \cdot 2^{(3(10\cdot 2^k)^2+2)\cdot k} 
    ~ = ~ 2^{-2^{11k} + 300k\cdot 2^{2k} +2k}
    ~ < ~ \frac12 \, .
  \]
  In other words, we have a $[0,n_k,s_k,c_k,\eps_k]$-protocol with $n_k
  > 1$, $c_k = o(\sqrt n_k)$ and $\eps_k < \frac12$. This contradicts
  Lemma~\ref{lem:zero-round}.

  Therefore, there must exist an $i$ such that $s_i \ge n_i/20$. Since
  $\ang{s_i}_{i=1}^k$ is increasing and $\ang{n_i}_{i=1}^k$ is
  decreasing, $s_k \ge n_k/20$. By the above calculations, $(48 \ln 2)2^{15k^2+11k}s
  \ge n/(20\cdot 3^k)$, which implies $s \ge n/2^{O(k^2)}$, as claimed.
\end{proof}

Notice that, for constant $k$, the argument in the above proof in fact
implies a lower bound for deterministic protocols with small enough
constant distributional error under $\mu_{c,n}$. This, in turn, extends
to distributional error under the uniform distribution, as remarked
earlier.


\section{Tight Deterministic One-Way Bounds}

The main result of this section is the following.

\begin{theorem}\label{thm:doneway}
  $D^\rightarrow(\ghd_{c,n}) = n - \Theta(\sqrt{n} \log n)$ for all
  constant $c$.
\end{theorem}

\begin{definition}
  Let $x_1,x_2,y \in \{0,1\}^n$.  We say that $y$ \emph{witnesses}
  $x_1$ and $x_2$ or that $y$ is a witness for $(x_1,x_2)$ if
  $x_1 \not\orth_c y$, $x_2 \not\orth_c y$, and $\ghd_{c,n}(x_1,y) \neq
  \ghd_{c,n}(x_2,y)$.
\end{definition}

Intuitively, if $(x_1,x_2)$ have a witness, then they cannot be in the
same message set.  For if Alice sent the same message on $x_1$ and $x_2$ and
Bob's input $y$ was a witness for $(x_1,x_2)$ then whatever Bob were to output,
the protocol would err on either $(x_1,y)$ or $(x_2,y)$.  The next
lemma characterizes which $(x_1,x_2)$ pairs have witnesses.

\begin{lemma}\label{lem:witness}
  For all $x_1,x_2 \in \{0,1\}^n$, there exists $y$ that witnesses
  $(x_1,x_2)$ if and only if $\dist(x_1,x_2) \geq 2c\sqrt{n}$.
\end{lemma}
\begin{proof}
  On the one hand, suppose $y$ witnesses $(x_1,x_2)$.  Then assume WLOG
  that $\dist(x_1,y) \leq n/2 - c\sqrt{n}$ and $\dist(x_2,y) \geq n/2 +
  c\sqrt{n}$.  By the triangle inequality, $\dist(x_1,x_2) \geq
  \dist(x_2,y)-\dist(x_1,y) = 2c\sqrt{n}$.  Conversely, suppose
  $\dist(x_1,x_2) \geq 2c\sqrt{n}$.  Let $L = \{i:x_1[i] = x_2[i]\}$,
  and let $R = \{i:x_1[i] \neq x_2[i]\}$.  Suppose $y$ agrees with $x_1$
  on all coordinates from $R$ and half the coordinates from $L$.  Then,
  $\dist(x_1,y) = |L|/2 = (n-\dist(x_1,x_2))/2 \leq n/2 - c\sqrt{n}$.
  Furthermore, $y$ agrees with $x_2$ on \emph{no} coordinates from $R$
  and half the coordinates from $L$, so $\dist(x_1,y) = |L|/2 + |R| \geq
  n/2 + c\sqrt{n}$.
\end{proof}

We show that it is both necessary and sufficient for Alice to send
different messages on $x_1$ and $x_2$ whenever $\dist(x_1,x_2)$ is
``large''.  To prove this, we need the following theorem, due to
Bezrukov~\cite{Bezrukov87} and a claim that is easily proved using the
probabilistic method (a full proof of the claim appears in the
appendix).

\begin{theorem}\label{thm:dmaximal}
  Call a subset $A \ceq \{0,1\}^n$ $d$-maximal if it is largest,
  subject to the constraint that $\dist(x,y) \leq d$ for all $x,y \in A$.
  \begin{enumerate}
  \item
    If $d =2t$ then $\cB(x,t)$ is $d$-maximal for any $x \in \{0,1\}^n$.
  \item
    If $d = 2t+1$ then $\cB(x,t) \cup \cB(y,t)$ is $d$-maximal for any
    $x,y \in \{0,1\}^n$ such that $\dist(x,y) = 1$. \qed
  \end{enumerate}
\end{theorem}

\begin{claim}\label{claim:owub}
  It is possible to cover $\{0,1\}^n$ with at most $2^{n -
  O(\sqrt{n}\log n)}$ Hamming balls, each of radius $c\sqrt{n}$.
  \qed
\end{claim}

\begin{proof}[Proof of Theorem~\ref{thm:doneway}]
  For the lower bound, suppose for the sake of contradiction that
  there is a protocol where Alice sends only $n - c\sqrt{n}\log n$
  bits.  By the pigeonhole principle, there exists a set $M \subseteq
  \{0,1\}^n$ of inputs of size $|M| \geq 2^n/2^{n-c\sqrt{n}\log n} =
  2^{c\sqrt{n}\log n} = n^{c\sqrt{n}}$ upon which Alice sends the same
  message.  By Theorem \ref{thm:dmaximal}, the Hamming ball
  $\cB(x,c\sqrt{n})$ is $2c\sqrt{n}$-maximal, and by Fact
  \ref{fact:smallball}, $|\cB(x,c\sqrt{n})| < |M|$.  Therefore, there must
  be $x_1,x_2 \in M$ with $\dist(x_1,x_2) > 2c\sqrt{n}$.  By Lemma
  \ref{lem:witness}, there exists a $y$ that witnesses $(x_1,x_2)$.  No
  matter what Bob outputs, the protocol errs on either $(x_1,y)$ or on
  $(x_2,y)$.

  For a matching upper bound, Alice and Bob fix a covering $\cC =
  \{\cB(x_0,r)\}$ of $\{0,1\}^n$ by Hamming balls of radius $r =
  c\sqrt{n}$.  On input $x$, Alice sends Bob the Hamming ball
  $\cB(x_0,r)$ containing $x$.  Bob selects some $x^\prime \in
  \cB(x_0,r)$ such that $x' \not\orth_c y$ and outputs
  $\ghd(x^\prime,y)$.  The correctness of this protocol follows from
  Lemma~\ref{lem:witness}, as $\dist(x,x^\prime) \leq 2c\sqrt{n}$ since
  they are both in $\cB(x_0,c\sqrt{n})$.  The cost of the protocol is
  given by Claim~\ref{claim:owub}, which shows that it suffices for
  Alice to send $\log\big(2^{n - O(\sqrt{n}\log n)}\big) = n -
  O(\sqrt{n}\log n)$ bits to describe each Hamming ball.
\end{proof}


\section{One Round Randomized Lower Bound} \label{sec:one-round}

Next, we develop a one-way lower bound for randomized protocols.  Note
that our lower bound applies to the uniform distribution, which, as
mentioned in Section~\ref{sec:intro}, implies space lower bounds for
the Distinct Elements problem under weaker assumptions about the input
stream.  Woodruff~\cite{Woodruff09} recently proved similar results,
also for the uniform distribution.  We include our lower bound as a
natural extension of the deterministic bound.

\begin{theorem}\label{thm:oneround}
  $R_{\eps}^\rightarrow(\ghd_{c,n}) = \Omega(n)$.
\end{theorem}
\begin{proof}
  For the sake of clarity, fix $c = 2$ and $\eps = 1/10$, and suppose
  $\cP$ is a one-round, $\eps$-error, $o(n)$-bit protocol for $\ghd_{c,n}$.

  \begin{definition}
    For $x \in \{0,1\}^n$, let $Y_x \deq \{y : x \not\orth_2 y\}$.  Say
    that $x$ is \emph{good} if $\Pr_{y \in Y_x}[\cP(x,y) = \ghd(x,y)]
    \leq 2\eps$.  Otherwise, call $x$ \emph{bad}.
  \end{definition}

  By Markov's inequality, at most a $1/2$-fraction of $x$ are
  \emph{bad}.  Next, fix Alice's message $\mathfrak{m}$ to maximize the
  number of \emph{good} $x$, and let $M = \{x\in\b^n:\, x \text{~is good
  and Alice sends~} \mathfrak{m} \text{~on input~} x\}$.  It follows
  that \[|M| \geq 2^{n-1}/2^{o(n)} > 2^{n(1-o(1))}.\] Our goal is to
  show that since $|M|$ is large, we must err on a $>2\eps$-fraction of
  $y \in Y_x$ for some $x \in M$, contradicting the goodness of $x$.
  Note that it suffices to show that a $4\eps$ fraction of $y \in
  Y_{x_1}$ witness $x_1$ and $x_2$.

  $|M| \geq 2^{n(1-o(1))}$, so by Fact~\ref{fact:largeball} and
  Theorem~\ref{thm:dmaximal}, There exist $x_1,x_2$ with $\dist(x_1,x_2) \geq
  1-o(1)$.  Next, we'd like to determine the probability that a random
  $y \in Y_{x_1}$ witnesses $(x_1,x_2)$.  Without loss of generality, let
  $x_1 = 0^n$.  Let $w(x) \deq \Pr_{y \in Y_{x_1}}[\ghd(x,y) \neq
    \ghd(x_1,y)]$.  The following lemma shows that $w(x)$ is an
  increasing function of $|x|$.  We leave the proof until the appendix.

  \begin{lemma}\label{lem:w}
    For all $x,x^\prime \in \{0,1\}^n$, $w(x) \geq w(x^\prime)
    \Leftrightarrow |x| \geq |x^\prime|$, with equality if and only if
    $|x| = |x^\prime|$.
  \end{lemma}

  We compute $w(x)$ by
  conditioning on $|y|$: 
  \[
    w(x) ~ = ~ \sum_{n_1 \leq n/2 - c\sqrt{n}}
    \Pr\left[\dist(x,y) \geq n/2 + c\sqrt{n} |\ |y| = n_1\right] 
    \cdot \Pr[|y| = n_1] \, .
  \]

  Fix $|x| =: m$, pick a random $y$ with $|y| = n_1$, and suppose
  there are $k$ coordinates $i$ such that $x_i = y_i$.  Then,
  $\dist(x,y) = (m-k) + (n_1-k) = m+n_1 - 2k$.  Hence, 
  \[
    \dist(x,y) \geq n/2 + c\sqrt{n} 
    ~\Longleftrightarrow~
    k \leq \frac{m+n_1}{2} - \frac{n}{4} - \frac{c}{2}\sqrt{n} \, .
  \]
  Note that given a random $y$
  with weight $|y| = n_1$, the probability that exactly $k$ of $m$
  coordinates have $x_i = y_i = 1$ follows the hypergeometric
  distribution $\Hyp(k; n,m,n_1)$.  Therefore, we can express the probability
  $\Pr_{|y| = n_1}[\dist(x,y) \geq n/2 + c\sqrt{n}]$ as 
  \[
    \Pr_{|y| = n_1}\left[ \dist(x,y) \geq n/2 + c\sqrt{n} \right] 
    ~ = ~ \sum_{k \leq \frac{m+n_1}{2} - \frac{n}{4} - \frac{c}{2}\sqrt{n}}
    \Hyp(k;n,m,n_1) \, .
  \]

  Finally, we show that $w(x) > 4\eps$ for a suitably large
  constant $|x|$ with the following claims, whose proofs are left to the
  appendix.

  \begin{claim}\label{claim:tooclose}
    Conditioned on $|y| \leq n/2 - 2\sqrt{n}$, we have
    $\Pr[|y| \geq n/2 - 2.1\sqrt{n}] \leq \frac13$.
  \end{claim}
  \begin{claim}\label{claim:hyp}
    For all $d < n/2 - 2.1\sqrt{n}$, 
    we have $\Pr[\dist(x_2,y) \geq n/2 + d\sqrt{n}]
    \geq 0.95$.
  \end{claim}

  Its easy to see from the previous two claims that $w(x) > 0.95\cdot(2/3) >4\eps$.
  %
  %
\end{proof}


\section{Concluding Remarks}

Our most important contribution here was to prove a multi-round lower
bound on a fundamental problem in communication complexity, the
Gap-Hamming Distance problem. As a consequence, we extended several
known $\Omega(1/\eps^2)$-type space bounds for various data stream
problems, such as the Distinct Elements problem, to multi-pass
algorithms. These resolve long-standing open questions. 

The most immediate open problem suggested by our work is to resolve
Conjecture~\ref{conj:ghd}. It appears that proving the conjecture true
is going to require a technique other than round elimination, or else,
an {\em extremely} powerful round elimination lemma that does not lose a
constant fraction of the input length at each step. On the other hand,
proving the conjecture false is also of great interest, and such a proof
might extend to nontrivial data stream algorithms, albeit with a
super-constant number of passes.

\section*{Acknowledgements}

We would like to thank Anna Gal, T.~S.~Jayram and David Woodruff for
stimulating discussions about the problem at various points of time.

{\small
  \bibliographystyle{alpha}
  \bibliography{../super}
}

\section*{APPENDIX}


\appendix

\section{Proofs of Technical Lemmas}

We begin with a proof of Claim~\ref{claim:owub}, which we state here for convenience.
\begin{claim}[Restatement of Claim~\ref{claim:owub}]
  For any constant $c$, it is possible to cover $\{0,1\}^n$ with at
  most $2^{n - O(\sqrt{n}\log n)}$ Hamming balls, each with radius $r
  = c\sqrt{n}$.
\end{claim}
\begin{proof}
  We use the probabilistic method.  Let $r \deq c \sqrt{n}$.  For $x
  \in \{0,1\}^n$, let $\cB_x \deq \cB(x,r)$ be the Hamming ball of
  radius $r$ centered at $x$.  For a $t$ to be determined later, pick
  $x_1,\ldots, x_t$ independently and uniformly at random from
  $\{0,1\}^n$.  We want to show that with nonzero probability, the
  universe $\{0,1\}^n$ is covered by these $t$ Hamming balls
  $\cB_{x_1}, \ldots, \cB_{x_t}$.

  Now, fix any $x \in \{0,1\}^n$ and any $1\leq i \leq t$.  Since
  $x_i$ was picked uniformly at random, each $x$ is equally likely
  to be in $\cB_{x_i}$.  Therefore, \[\Pr[x \in \cB_{x_i}] =
  \frac{|\cB_{x_i}|}{2^n} \geq 2^{\theta(\sqrt{n}\log n) - n}\] where
  inequality stems from Fact~\ref{fact:smallball}.

  Let $BAD_x = \bigwedge_{1 \leq i\leq t} x \not\in \cB_{x_i}$ be the
  event that $x$ is not covered by any of the Hamming balls we picked
  at random, and let $BAD = \bigvee BAD_x$ be the event that
  \emph{some} $x$ is not covered by the Hamming balls.  We want to
  limit $\Pr[BAD]$.  $BAD_x$ occurs when $x \not\in \cB_{x_i}$ for all
  $x_i$.  Therefore, using $1 - x \leq e^{-x}$ for all real
  $x$, \[\Pr[BAD_x] = \left( 1 - 2^{\theta(\sqrt{n}\log n) -
    n}\right)^t \leq \e^{-t\cdot 2^{\theta(\sqrt{n}\log n)-n}}.\] By
  the union bound, \[\Pr[BAD] \leq 2^n \Pr[BAD_x] = 2^{n -
    \frac{t}{\ln 2}2^{\theta(n \sqrt{n}) - n}}.\] Picking $t = \ln
  2(n+1)2^{n - \theta(\sqrt{n}\log n)} = 2^{n-\theta(\sqrt{n}\log n)}$
  ensures that $\Pr[BAD] < 1$.  Therefore, there exists a set of $t =
  2^{n-\theta(\sqrt{n}\log n)}$ Hamming balls of radius $c\sqrt{n}$
  that cover $\{0,1\}^n$.
\end{proof}

Recall that $w(x) = \Pr_{y \in Y_{\vec{0}}}[\ghd(x,y) \neq
  \ghd(\vec{0},y)]$.
\begin{lemma}[Restatement of Lemma~\ref{lem:w}]
  For all $x,x^\prime \in \{0,1\}^n$, $w(x) \leq w(x^\prime)$ if and
  only if $|x| \leq |x^\prime|$, with equality if and only if $|x| =
  |x^\prime|$.
\end{lemma}
\begin{proof}
  If $|x| = |x^\prime|$, then $w(x) = w(x^\prime)$ by symmetry.  
  Further, note that $\ghd(x,y) = 0$ if and only if $\ghd(-x,y) = 1$.
  Therefore, it suffices to handle the case where $|y| \leq n/2 -
  c\sqrt{n}$ and $\ghd(\vec{0},y) = 0$.  

  For the rest of the proof, we assume that $x_i = x_i^\prime$, except
  for the $nth$ coordinate, where $x_n = 0$ and $x_n^\prime = 1$.
  Thus, $|x| = |x^\prime| - 1$.  We show that $w(x) < w(x^\prime)$; the
  rest of the lemma follows by induction.

  Let $Y$ be the set of strings with Hamming weight $|y| \leq n/2 -
  c\sqrt{n}$.  Partition $Y$ into the following three sets:
  \begin{itemize}
  \item $A \deq \{y : |y| = n/2 + c\sqrt{n} \wedge y_n = 0\}$.
  \item $B \deq \{y : |y| < n/2 + c\sqrt{n} \wedge y_n = 0\}$.
  \item $C \deq \{y : y_n = 1\}$.
  \end{itemize}
  Note the one-to-one correspondence between strings in $B$ and
  strings in $C$ obtained by flipping the $nth$ bit.  Now, consider
  any $y \in B$ such that $y$ witnesses $(\vec{0}, x^\prime)$ but not
  $(\vec{0},x)$.  Flipping the $nth$ bit of $y$ yields a string
  $y^\prime \in C$ such that $Y$ witnesses $(\vec{0},x)$ but not
  $(\vec{0},x^\prime)$.  Hence among $y \in B\cup C$ there is an equal
  number of witnesses for $x$ and $x^\prime$.  For any $y \in A$, $y_n
  = 0$, whence $|y-x^\prime| = |y-x| + 1$.  Therefore, any $y$ that
  witnesses $(\vec{0},x)$ must also witness $(\vec{0},x^\prime)$,
  whence $w(x) \leq w(x^\prime)$.
\end{proof}

Many claims in this paper require tight upper and lower tail bounds
for binomial and hypergeometric distributions.  We use Chernoff bounds
where they apply.  For other bounds, we approximate using normal
distributions.  We use Feller~\cite{Feller-book} as a reference.

\begin{definition}
  For $x \in \R,$ let $\phi(x) \deq e^{-x^2/2}/\sqrt{2\pi}$
  and \[N(x) \deq \int_x^{\infty} \phi(y)dy.\]
\end{definition}

$N(x)$ is the cumulative distribution function of the normal
distribution.  We use it in Fact~\ref{fct:normal-tail} to approximate
$T(x)$.  Here, we'll also use it to approximate tails of the binomial
and hypergeometric distributions.

\begin{lemma}[Feller, Chapter VII, Lemma 2.]\label{lem:feller7lem2}
  For all $x > 0$, \[\phi(x)\left(\frac{1}{x} - \frac{1}{x^3}\right) <
  N(x) < \phi(x)\frac{1}{x}.\]
\end{lemma}


\begin{theorem}[Feller, CHapter VII, Theorem 2.]\label{thm:feller7th2}
  For fixed $z_1,z_2$, \[Pr[n/2 + (z_1/2)\sqrt{n} \leq |y| \leq n/2 +
    (z_2/2)\sqrt{n}] \sim N(z_1) - N(z_2).\]
\end{theorem}

\begin{theorem}
  For any $\gamma$ such that $\gamma = \omega(1)$ and $\gamma =
  o(n^{1/6})$, we have \[\sum_{k > n/2 + \gamma \sqrt{n}/2}
  \binom{n}{k} \sim N(\gamma).\]
\end{theorem}

\begin{claim}[Restatement of Claim~\ref{claim:tooclose}]
        Conditioned on $|y| \leq n/2 - 2\sqrt{n}$, 
  \[Pr[|y| \geq n/2 - 2.1\sqrt{n}] \leq 1/3.\]
\end{claim}
\begin{proof}
  By Theorem~\ref{thm:feller7th2} and Lemma~\ref{lem:feller7lem2}, we have
  \begin{eqnarray*}
    \Pr[n/2 - 2.1\sqrt{n} \leq |y| \leq n/2 - 2\sqrt{n}] &\sim& N(4)-N(4.2) \\
    &\leq& \phi(4)/4 - \phi(4.2)(4.2^{-1} - 4.2^{-3}) \\
    &\leq& 2.0219*10^{-5}
  \end{eqnarray*}
  By Fact\ref{fct:normal-tail}, $\Pr[|y| \geq n/2 - 2\sqrt{n}] \leq
  2^{-3\cdot2^2-2} = 2^{-14} = 6.1035\cdot10^{-5}$.  Putting the two terms together, we get
  \[\Pr[|y| \geq n/2 - 2.1\sqrt{n} | |y| \leq n/2 - 2\sqrt{n}] \leq \frac{2.0219\cdot10^{-5}} {6.1035\cdot10^{-5}} \leq 1/3.\]
\end{proof}
\begin{claim}[Restatement of Claim~\ref{claim:hyp}]
  For all $d < n/2 - 2.1\sqrt{n}$, \[\Pr[\dist(x_2,y) \geq n/2 + 2\sqrt{n}]
  \geq 0.95.\]
\end{claim}
\begin{proof}
  The proof follows from the following claim, instantiated with $c = 2$ and $\alpha = 2.1$.
\end{proof}
%
%
\begin{claim}
  For all $\alpha > c$, $|x| = \gamma n$, and all $\gamma \geq 1 -
  (1-c/\alpha)/4$, \[\Pr_{|y| = n/2 - \alpha \sqrt{n}}[\dist(x,y) \geq n/2
    + c\sqrt{n}] \geq 1-\exp\left(-\frac{2(\alpha
      -c)\alpha^2(1+o(1))}{3\alpha + c}\right).\]
\end{claim}
\begin{proof}
  Let $m \deq |x| = \gamma n$ and let $n_1 = n/2 - \alpha \sqrt{n}$.
  Then, the probability that a random $y$ with $|y| = n_2$ can be
  expressed using the hypergeometric distribution $\Hyp(k; n,m,n_1)$.
  Let the $m$ set bits of $x$ be the defects.  The probability of $k$
  of the $n_1$ bits of $y$ are defective is $\Hyp(k;n,m,n_1)$.  Note
  that $\dist(x,y) = (m-k) + (n_1-k) = m+n_1 - 2k$.  Therefore, \[\dist(x,y)
  \geq n/2 + c\sqrt{n} \Leftrightarrow k \leq \frac{m+n_1}{2} -
  \frac{n}{4} - \frac{c}{2} \sqrt{n}. = \frac{\gamma n}{2} -
  \frac{\alpha +c}{2}\sqrt{n}\] We express the probability $\Pr_{|y| =
    n_1}[\dist(x,y) \geq n/2 + c\sqrt{n}]$ as \[\Pr_{|y| = n_1}[\dist(x,y) \geq
    n/2 + c\sqrt{n}] = \Pr_{K \sim \Hyp(k;n,m,n_1)}[K \leq \frac{\gamma
      n}{2} - \frac{\alpha +c}{2}\sqrt{n}].\] Next, we use a
  concentration of measure result due to Hush and Scovel~\cite{HushS05}.  Here, we
  present a simplified version.
  \begin{theorem}[Hush, Scovel]\label{thm:hypmeasure}
    Let $m = \gamma n > n_1 = n/2 - \alpha \sqrt{n}$, and let $\beta = n/m(n-m)$.
    \[\Pr[K - E[K] > \eta] < \exp(-2\beta\eta^2(1+o(1))).\]
  \end{theorem}

  The expected value of a random variable $K$ distributed according to
  $\Hyp(K;n,m,n_1)$ is \[E[K] = \frac{m n_1}{n} = \frac{\gamma
    n}{n}\left(\frac{n}{2} - \alpha \sqrt{n}\right) = \frac{\gamma
    n}{2} - \gamma \alpha \sqrt{n}.\] Set $\eta \deq (\alpha
  -c)\sqrt{n}/4$.  Note that \[E[K] + \eta = \frac{\gamma n}{2}
  -\gamma \alpha \sqrt{n} + \frac{\alpha - c}{4}\sqrt{n} \leq
  \frac{\gamma n}{2} - \frac{\alpha + c}{2}\sqrt{n} = \frac{m+n_1}{2}
  - \frac{n}{4} - \frac{c}{2}\sqrt{n}\] where the inequality holds
  because $\gamma \geq 1 - (1-c/\alpha)/4$.  Note also that
  $(1-c/\alpha)/4 = (\alpha -c)/4\alpha$, so $1-(1-c/\alpha)/4 =
  (3\alpha + c)/4\alpha$.  By Theorem~\ref{thm:hypmeasure}
  \begin{eqnarray*}
    \Pr[K > \frac{\gamma n}{2} - \frac{\alpha + c}{2}\sqrt{n}] &=&
    \Pr[K-E[K] > \eta] \\ &<&
    \exp\left(-\frac{2n\eta^2(1+o(1))}{m(n-m)}\right) \\ &=&
    \exp\left(-\frac{2(\alpha -
      c)^2(1+o(1))}{16\gamma(1-\gamma)}\right) \\ &\leq&
    \exp\left(-\frac{2(\alpha - c)^2(4\alpha)^2(1+o(1))}{16(\alpha -
      c)(3\alpha + c)}\right) \\ &=& \exp\left(-\frac{2(\alpha
      -c)\alpha^2(1+o(1))}{3\alpha + c}\right)
  \end{eqnarray*}
  It follows that $\Pr[K \leq \frac{\gamma n}{2} - \frac{\alpha +
      c}{2}\sqrt{n}] \geq 1-\exp\left(-\frac{2(\alpha
    -c)\alpha^2(1+o(1))}{3\alpha + c}\right)$.
\end{proof}
\begin{claim}
  For any $x_L \in \{0,1\}^{n_L}$, $\ghd(x_L,y_L)$ is defined for at
  least a $\geq e^{-2(c^\prime)^2}/5c^\prime$-fraction of $y_L \in
  \{0,1\}^{n_L}$.
\end{claim}
\begin{proof}
  Without loss of generality, assume $x_L = \vec{0}$.  Then,
  $\ghd(x_L,y_L)$ is defined for all $y$ such that $|y| \leq n_L/2 -
  c^\prime\sqrt{n_L}$ or $|y| \geq n_L/2 + c^\prime \sqrt{n_L}$.  Note
  that for any constant $x > c^\prime$, 
  \begin{eqnarray*}
    \Pr_{y}[|y| \leq \frac{n_L}{2} - c^\prime \sqrt{n_L}] &\geq&
    \Pr[\frac{n_L}{2} -x\sqrt{n_L} \leq |y| \leq \frac{n_L}{2} -
      c^\prime \sqrt{n_L}] \\ &\geq& N(2c^\prime) - N(2x) \\ &\geq&
    \phi(2c^\prime)\left(\frac{1}{2c^\prime} -
    \frac{1}{(2c^\prime)^3}\right) - \frac{\phi(2x)}{2x} \\ &=&
    \frac{e^{-(2c^\prime)^2/2}}{\sqrt{2\pi}}(\left(\frac{1}{2c^\prime}
    - \frac{1}{(2c^\prime)^3}\right) - \frac{e^{-2x^2}}{2x\sqrt{2\pi}}
    \\ &\geq& \frac{e^{-2(c^\prime)^2}}{10c^\prime}
  \end{eqnarray*}
  $\Pr[|y| \geq n_L/2 + c^\prime \sqrt{n_L}]$ is bounded in the same fashion.
\end{proof}

\end{document}